\documentclass[useAMS,usenatbib,usegraphicx]{mn2e}

\usepackage[english]{babel}
\usepackage[utf8]{inputenc}
\usepackage[none]{hyphenat}

\makeatletter
\def\@maketitle{\newpage
 \vspace*{7pt}
 {\raggedright \sloppy
  {\reset@font\huge \bf \@title \par}
  \vskip 23pt
  {\reset@font\LARGE
   \begin{tabular}[t]{@{}l@{}}\let\\=\author@nextline\@author
   \end{tabular}
   \par}
  \vskip 12pt
 }
 \par\noindent
 {\reset@font\small \@date \par}
 \vskip 12pt
}
\def\nokeywords{\ifSFB@keywords\else
 \if@twocolumn \start@SFBbox\addvspace{12pt}\finish@SFBbox \fi
 \@thanks
 \gdef\@thanks{}\fi
}
\def\section{%
 \if@firstsection \fixfootnotes\@firstsectionfalse \fi%
 \@startsection{section}{1}{\z@}
 {-6pt plus -12pt minus -1pt}{3pt}
 {\SFB@hangraggedright\reset@font\normalsize\bf}}
\def\fps@figure{ht}
\makeatother

\title[Dynamics of gas and dust clouds in AGN]{Dynamics of gas and dust clouds in active galactic nuclei}
\author[P.M. Plewa, M. Schartmann and A. Burkert]{P.M. Plewa$^{1,2,}$\thanks{E-mail: pmplewa@mpe.mpg.de}, M. Schartmann$^{1,2}$ and A. Burkert$^{1,2,}$\thanks{Max Planck Fellow},\\
$^{1}$Universit\"ats-Sternwarte M\"unchen, Scheinerstra\ss e 1, D-81679 M\"unchen, Germany\\
$^{2}$Max-Planck-Institut f\"ur extraterrestrische Physik, Postfach 1312, Giessenbachstra\ss e, D-85741 Garching, Germany}

\begin{document}
\label{firstpage}

\date{March 12, 2013}
\maketitle
\begin{abstract}
We analyse the motion of single optically thick clouds in the potential of a central mass under the influence of an anisotropic radiation field $\sim |\cos(\theta)|$, a model applicable to the inner region of active galactic nuclei. Resulting orbits are analytically soluble for constant cloud column densities. All stable orbits are closed, although they have non-trivial shapes. Furthermore, there exists a stability criterion in the form of a critical inclination, which depends on the luminosity of the central source and the column density of the cloud.
\end{abstract}
\nokeywords

\section{Introduction}
\label{sec:introduction}

Most -- if not all -- galaxies harbour supermassive black holes (SMBH) at their centres. Gas infall from larger scales leads to the formation of an accretion disc, where gravitational energy is dissipated to viscously heat the gas. The emerging radiation -- peaking in the ultraviolet~(UV)/optical wavelength regime -- illuminates the so-called Broad Line Region (BLR) on small scales as well as a large gas and dust reservoir on parsec scale distance from the SMBH. The latter was postulated in order to unify two classes of observed objects by means of an inclination effect, allowing only unobscured views of the central region for face-on orientation of this so-called dusty torus. This is the essence of the widely accepted Unified Scheme of Active Galactic Nuclei \citep[AGN,][]{Antonucci_93}. Recent observations and modelling efforts provide evidence that the internal structure of those tori is likely clumpy \citep[e.~g.~][]{Krolik_88,Nenkova_02,Hoenig_06,Tristram_07,Schartmann_08}. Successively improved radiative transfer simulations have been used to draw conclusions about the properties of individual clouds and the geometry of their distribution \citep[e.~g.~][]{Nenkova_08b,Hoenig_10b,Ramos_09}. Although differing in detail, various models are able to reproduce several types of observed spectral energy distributions and interferometric data successfully. However, the cloud distribution in these models is static and there is only little known about the dynamical properties of such cloud systems. Concerning the structure of the BLR, several observed occultation events suggest the existence of a clumpy structure as well \citep[e.~g.~][]{Risaliti_11}.

In both situations, radiative forces on either gas or dust clouds can be of similar strength as gravitational forces. The particular significance of anisotropic radiation pressure for the distribution of dusty gas has been pointed out by \citet{Liu_11}. In order to understand the dynamical stability properties of cloud orbits under the combined effects of strong gravity and radiation pressure, we start our analysis by investigating the dynamics of simplified single-cloud models. Many aspects of orbits in radiation fields have already been explored, for example by \citet{Saslaw_78} or \citet{Mioc_92}, who use a perturbative ansatz to examine orbits in general time-varying and anisotropic radiation fields, respectively. We aim to analyse particularly the orbits of gas and dust clouds in the central vicinity of AGN, simplified to a central mass and the constant radiation field of a surrounding accretion disc. In a similar setup \citet{Krause_11} investigated perturbations to circular orbits of pressure-confined BLR clouds. We derive the orbit equation analytically under the alternative assumption of constant cloud column density. This will enable us to constrain cloud parameters as well as the distribution of clouds in AGN tori or the BLR.

In Section~\ref{sec:forces} we calculate the effective force acting on a cloud and solve the equation of motion in Section~\ref{sec:orbits}. We proceed to analyse orbit stability in Section~\ref{sec:stability} and describe the characteristics of the orbit family in Section~\ref{sec:shapes}, which is followed by a critical discussion (Section~\ref{sec:discussion}) and the conclusions (Section~\ref{sec:conclusions}).

\section{Forces on optically thick clouds}
\label{sec:forces}

For the work presented in this Letter, we concentrate on the two dominant forces close to a black hole plus accretion disc system: 
the force $\bmath{F}_{grav}$ on a cloud of mass $m$ in the gravitational potential of a central point mass $M$ at distance $r$ from the cloud is given by Newton’s law:
\begin{equation}\bmath{F}_{grav}=-\frac{GMm}{r^2}\bmath{e}_r,\end{equation}
where $G$ is the gravitational constant and $\bmath{e}_r$ is the unit vector in radial direction. Often of similar dynamical importance is the radiation pressure exerted from a viscously heated accretion disc. Compared to the cloud distances of interest here, the accretion disc emits the bulk of its energy close to its inner edge and can hence be considered point-like. Assuming that the cloud is optically thick for the whole wavelength range where most of the energy is radiated, the additional force by radiation pressure can be written as
\begin{equation}\bmath{F}_{rad}=\frac{\sigma}{c}\mathcal{F}\bmath{e}_r\end{equation}
with the anisotropic radiation flux 
\begin{equation}\mathcal{F}(r,\theta)=\frac{L|\cos(\theta)|}{2\pi r^2},\end{equation}
where $\sigma$ is the cloud’s cross-section and $\theta$ the polar angle, measured from the normal to the radiating disc. The anisotropy arises from the change in projected surface area of the infinitesimally thin accretion disc of bolometric luminosity $L$ assumed here.
Thus, the cloud is moving in a force field (per unit of mass)
\begin{equation}\bmath{f}(r,\theta)=\frac{GM}{r^2}\left(\frac{3l}{\mu N_{cl}\sigma_T}|\cos(\theta)|-1\right)\bmath{e}_r,\end{equation}
where $\mu$ is the mean molecular weight and $\nobreak{N_{cl} = \frac{3}{2} m / \mu m_p \sigma}$ is the column density through the centre of a single spherical cloud, which is equivalent to an optical depth and assumed to be constant throughout the evolution. $l = L / L_\mathrm{edd}$ is the Eddington ratio where $L_\mathrm{edd} = 4 \pi G M m_p c / \sigma_\mathrm{T}$ is the Eddington luminosity and $\sigma_\mathrm{T}$ the Thomson cross-section. It is straightforward to see that -- in general -- the radiation field effectively changes the gravitational constant $G$ or the central mass $M$ continuously throughout the orbit. Angular momentum conservation keeps the orbital plane fixed in space and hence equatorial orbits are unaffected by radiation pressure. We discuss possible contributions from other forces like, e.~g.,~drag forces in Sect.~\ref{sec:discussion}

\section{Cloud orbits}
\label{sec:orbits}

Since the total force on a cloud is always radial, specific angular momentum $L$ is conserved and thus the orbit remains in a plane with fixed inclination $i$ against the accretion disc. It is sufficient to consider $0 \leq i \leq \pi / 2$. Let $\psi$ be the angle in this plane as measured from the ascending node, then $\cos(\theta) = \sin(i)\sin(\psi)$ and the radial equation of motion is
\begin{equation}\ddot{r}-\frac{L^2}{r^3}=\frac{GM}{r^2}\Bigl(k|\sin(\psi)|-1\Bigr),\end{equation}
where $L = r^2 \dot\psi$ and 
\begin{equation}k=\frac{3l}{\mu N_{cl}\sigma_T}\sin(i).\label{eq:k}\end{equation}
Changing the independent variable from time to angle $\psi$ and substituting $u=1/r$, the equation of motion can be written as
\begin{equation}\frac{d^2u}{d\psi^2}+u=\frac{GM}{L^2}\Bigl(1-k|\sin(\psi)|\Bigr).\end{equation}
This differential equation has the solution
\begin{eqnarray}\label{eq:orbits1}
u(\psi)=\frac{GM}{L^2}\Bigl[1+\frac{k}{2}\psi\cos(\psi)+\alpha\sin(\psi)+\beta\cos(\psi)\Bigr]
\end{eqnarray}
for $0\leq\psi\leq\pi$ and similarly for $\pi<\psi\leq2\pi$,
\begin{eqnarray}\label{eq:orbits2}
u(\psi)&=&\frac{GM}{L^2}\Bigl[1-\frac{k}{2}\psi\cos(\psi)+(\alpha+k)\sin(\psi)+ \\
\nonumber &+&(\beta+k\pi)\cos(\psi)\Bigr]
\end{eqnarray}
with arbitrary constants $\alpha$, $\beta$ and under the condition that both $u$ and $du / d\psi$ are continuous at $\psi = \pi$. The parameters $M$ and $L$ appear as scaling factors and have no influence on neither stability nor shape of the orbit.

\section{Stability of orbits}
\label{sec:stability}

The orbit equations~\ref{eq:orbits1} and~\ref{eq:orbits2} allow for unbound orbits in such way that $u$ can have a root on $0\leq\psi\leq 2\pi$. Requesting $u$ to be at least positive at $\psi=n\cdot\pi/2$ for all integer $n$, results in the two constrains $\nobreak{-1 < \alpha < 1 - k}$ and $\nobreak{-1 < \beta < 1 - k \pi / 2}$. The latter hints that there is a critical value of $k_c = 4 / \pi$, for which bound orbits become impossible. This has been confirmed numerically by solving the respective orbits for an extensive number of parameter combinations. Thus, equation~\ref{eq:k} implies a critical inclination for any particular choice of column density and luminosity, given by
\begin{equation}i_c=\arcsin\left(\frac{4}{\pi}\frac{\mu N_{cl}\sigma_T}{3l}\right)\end{equation}
above which all orbits are unstable.

\begin{figure}
  \includegraphics[width=\linewidth]{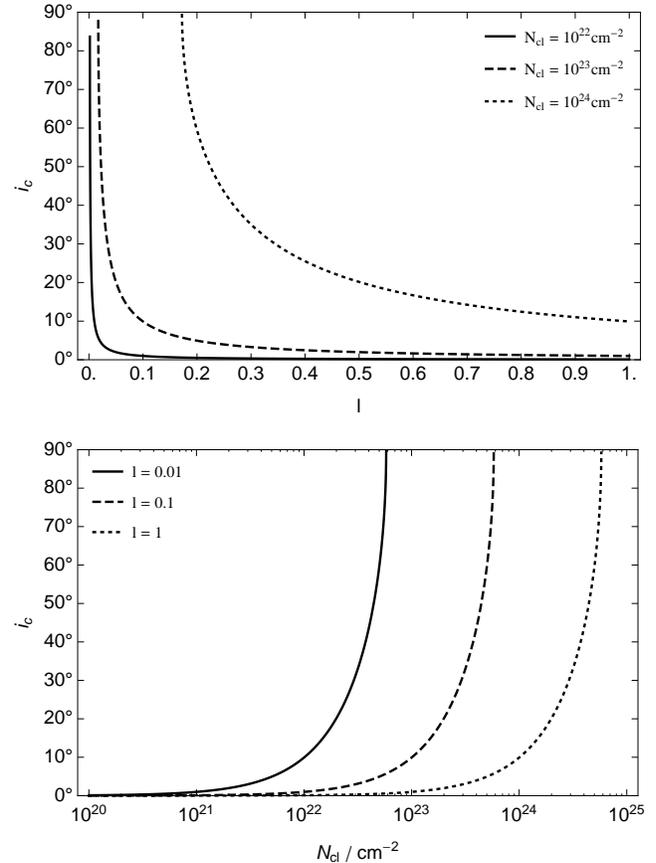}
  \caption{Critical orbit inclination vs. Eddington ratio of the central source and cloud column density ($\mu = 0.61$).}
  \label{fig:i}
\end{figure}

Some typical values are illustrated in Fig.~\ref{fig:i}, where $\nobreak{\mu = 0.61}$ is chosen representative for solar metallicity gas. For example, in a typical Seyfert galaxy with Eddington ratio of 0.1, a cloud with column density $10^{23}$cm$^{-2}$ is restricted to angles smaller than $10^\circ$. If the column density is greater by a factor of about 6, polar orbits become allowed. A similar argument can be made if the column density is kept fixed, namely that the greater the Eddington ratio of the source, the smaller the region where stable orbits are possible.
\begin{figure}
  \includegraphics[width=\linewidth]{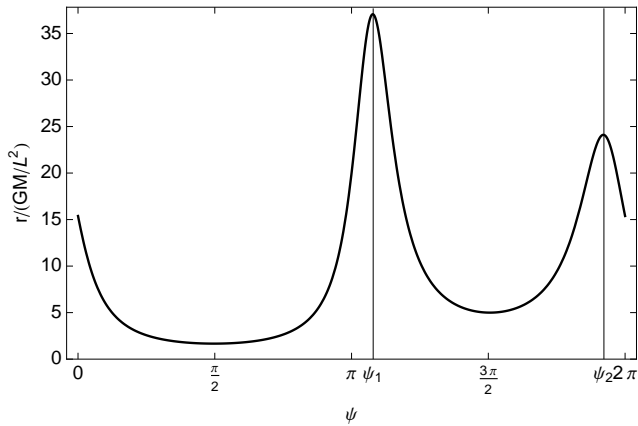}
  \caption{Distance of the cloud to the central mass over one orbit for case (e) from Fig.~\ref{fig:orbit}.}
  \label{fig:r}
\end{figure}

\newpage\noindent Because $k_c > 1$, the critical inclination is always larger than the inclination defined by force equilibrium ($\bmath{f} = 0$), up to a factor of $\sim 1.74$. This is plausible, because clouds on orbits with a comparatively high inclination spend only a relatively short amount of time in the radiation-dominated region.

\section{Shape of stable orbits}
\label{sec:shapes}

It can be verified that all bound orbits are closed by evaluating $u$ and $du / d\psi$ at $\psi = \epsilon$ from equation~\ref{eq:orbits1} and at $\psi = 2 \pi - \epsilon$ from equation~\ref{eq:orbits2} and letting $\epsilon$ go to zero. That is to say the total energy averaged over one orbit is constant, although the force field is non-conservative. Closed orbits are not expected under different assumptions, such as variable $N_{cl}$ or $l$, which might be expected for a pressure-confined cloud with constant mass adjusting instantly to pressure equilibrium within a stratified atmosphere.

The family of stable orbits features a range of different orbit shapes. For small values of $k$, orbits are close to elliptical and reduce to Kepler orbits if $k$ approaches zero. This is the case for either zero inclination, zero luminosity or very high column density. If the value of $k$ is close to critical, all stable orbits are eccentric, show a waist and can also be asymmetric. There are then two local maxima of the cloud's distance to the central mass over one orbit, as shown for example in Fig.~\ref{fig:r}. In addition to the eccentricity $e$ calculated from apocentre and pericentre of the orbit, it is convenient to use another similarly defined quantity
\begin{equation}e'=\frac{r(\psi_1)-r(\psi_2)}{r(\psi_1)+r(\psi_2)}\end{equation}
calculated from these two maxima, together with the angular offset $\Delta\psi = \psi_2 - \psi_1$ between them, to distinguish between characteristic shapes.
\begin{figure}
  \includegraphics[width=\linewidth]{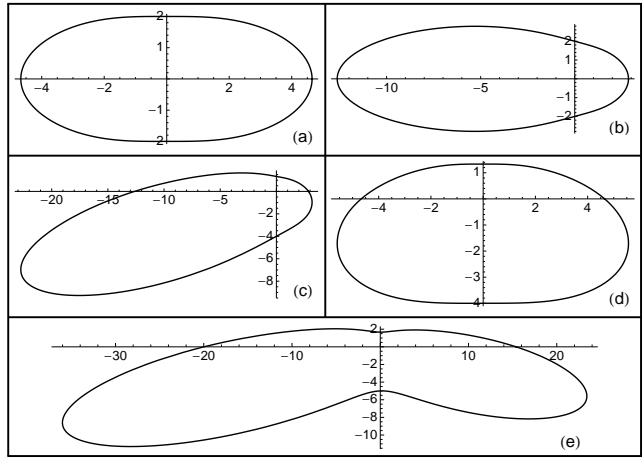}
  \caption{Characteristic shapes of orbits depicted in the orbital plane according to equations~\ref{eq:orbits1} and~\ref{eq:orbits2}. Distances are marked in units of $G M / L^2$. The parameters for each orbit can be found in Table~\ref{tab:param}.}
  \label{fig:orbit}
\end{figure}
\begin{table}
\centering
\begin{minipage}{0.85\linewidth}
  \caption{Parameters for the example orbits in Fig.~\ref{fig:orbit} illustrated in section~\ref{sec:shapes}.}
  \label{tab:param}
  \begin{tabular}{ccccccc}
    \hline
    \  & $k$ & $\alpha$ & $\beta$ & $e$ & $e'$ & $\Delta\psi-\pi$\\
    \hline
    (a) & 1 & -0.5 & -0.785 & 0.40 & 0 & 0\\
    \hline
    (b) & 1 & -0.5 & -0.65 & 0.73 & 0.63 & 0\\
    \hline
    (c) & 1 & -0.25 & -0.65 & 0.90 & 0.75 & 0.32 $\pi$\\
    \hline
    (d) & 1 & -0.25 & -0.785 & 0.63 & 0 & 0.26 $\pi$\\
    \hline
    (e) & 1.2 & -0.4 & -0.935 & 0.91 & 0.21 & 0.16 $\pi$\\
    \hline
    \end{tabular}
\end{minipage}
\end{table}

Some exemplary orbits are shown in Fig.~\ref{fig:orbit} with parameters summarized in Table~\ref{tab:param}. The transition from orbit~(a) to~(e), which have identical $k$, can be understood in such way that the near elliptical orbit~(a) is the most circular orbit possible in this case. Orbit~(b) has two distinguishable apocentres ($e' \neq 0$) and shows a pronounced waist, but the apocentres still lie opposite to each other ($\Delta\psi = \pi$). In contrast to that orbit~(c) is completely asymmetric, which is the most prevalent shape. There is again an axis of symmetry for orbit~(d) because here $e'$ vanishes, although $\Delta\psi \neq \pi$. Orbit~(e) is of similar type as orbit~(c), but for a slightly higher $k$, which emphasizes the above-mentioned waist.

\section{Discussion}
\label{sec:discussion}

This Letter represents a first step towards understanding the dynamical properties of optically thick clouds in point potentials including an anisotropic radiation source. Such simulations and theoretical predictions can be used to assess the dynamical structure as well as the stability of AGN tori and the BLR.

For a general treatment of the problem, several simplifications are necessary. First of all, we assume that the clouds move through vacuum. In reality, e.~g.,~AGN tori are thought to be made up of a multiphase medium, where dense and cold clouds are embedded in a hot and diffuse ionized gas component. This intercloud medium leads to ram pressure interaction with the cloud, which has two effects: (i)~locally it compresses the front part of the cloud and (ii)~globally it leads to a deceleration and thereby a change of the orbital properties and finally an inspiral of the cloud towards the centre. Related is the issue of confinement of the clouds, as hydro instabilities (e.~g.~\citealp{Cowie_77}), radiation pressure (e.~g.~\citealp{Schartmann_11}), thermal conductivity (e.~g.~\citealp{McKee_77,Burkert_12,Schartmann_12}) or collisions with other clouds \citep{Krolik_88} would lead to the dissolution of the clouds on time-scales of the order of the dynamical time or even faster, depending on the geometry of the cloud distribution. An additional issue is the built-up of a dense underlying disc through dissipative collisions, which would eventually lead to a collapse of the cloud distribution. Such equatorial discs are indeed found in coexistence with geometrically thick AGN tori in nearby Seyfert nuclei, probed with the help of near-infrared interferometry as well as in maser emission \citep{Tristram_07,Raban_09,Schartmann_10}. Sufficiently elastic cloud-cloud collisions (e.~g.~due to strong internal magnetic fields) would enable the exchange of energy and thereby alter the initial orbital structure as well, leading at the same time to an accretion flow towards the centre and an increase of the velocity dispersion of the cloud ensemble \citep{Krolik_88,Vollmer_04}. Depending on the confinement, clouds might change their size when, e.~g.,~moving through a stratified atmosphere which then in turn affects the radiation pressure force and hence the orbital evolution \citep[e.~g.~][]{Netzer_10,Krause_11}. In addition magnetic fields influence the dynamical evolution and might contribute either to a faster destruction or to long lasting confinement of clouds \citep{Krause_12}, depending on the -- generally unknown -- magnetic field topology.

\section{Conclusions}
\label{sec:conclusions}

The shape and stability of an optically thick cloud’s orbit around a central black hole with a surrounding accretion disc are strongly affected by radiation pressure. Although deviating from ellipses, we show that all bound orbits are closed. The stability of an orbit is determined by the ratio of the disc's Eddington fraction to the cloud's column density and in particular the inclination of the orbit.

This has a possible application in the study of AGN tori, which are composed of dusty clouds of gas. The clouds can reasonably be assumed to have a high optical thickness because dust absorption and the emitted spectral energy distribution of the central engine both peak in the UV regime. Therefore, radiation pressure is likely to influence the dynamics of clumpy AGN tori significantly. Compared to the derived limit on the inclination of stable cloud orbits, the geometrical thickness of a real torus might be amplified due to an outflow consisting of ejected clouds (\citealp{Emmering_92,Konigl_94,Elitzur_06}, and references therein). These could originate from a distribution with large covering angle at the onset of activity or from an accretion flow triggered by cloud-cloud collisions. This outflow would be of particular importance for large ratios $l / N_{cl}$, e.~g.~in Quasars. When supply for the outflow has ceased, there could still be left a rather thin but long-lived torus.

However, for a full treatment of cloud ensembles, the interaction of multiple clouds would need to be considered, such as fluctuating radiation pressure caused by shadowing of the central light source, secondary radiation pressure from heated dust and merging or collisions of clouds. In this environment, magnetic fields might be able to support clouds against destruction by tidal or hydrodynamic forces to ensure the stability of clouds on the time-scale of one orbit and also provide the necessary coupling of the dust and gas components.

\section*{Acknowledgements}

We are grateful to Scott Tremaine for most helpful discussions and thank the anonymous referee for comments to improve the Letter.

\label{lastpage}
\end{document}